\documentclass[10pt,aps,prb,twocolumn,superscriptaddress,floatfix]{revtex4-2}

\usepackage{amsmath,amssymb,physics}
\usepackage{graphicx}
\usepackage{dcolumn}
\usepackage{bm}
\usepackage{xcolor}
\begin{document}

\title{Tomography of Transport Pathways in Selective-Area-Grown Nanowires Using Angle-Resolved Conductance Fluctuations}

\author{Christian E. N. Petersen}
\email{cenpe@dtu.dk}

\affiliation{
Department of Energy Conversion and Storage,
Technical University of Denmark,
2800 Kgs. Lyngby, Denmark
}

\author{Damon J. Carrad}
\affiliation{
Department of Energy Conversion and Storage,
Technical University of Denmark,
2800 Kgs. Lyngby, Denmark
}
\author{Daria Beznasyuk}
\affiliation{
Department of Energy Conversion and Storage,
Technical University of Denmark,
2800 Kgs. Lyngby, Denmark
}
\author{Thomas S. Jespersen}
\affiliation{
Department of Energy Conversion and Storage,
Technical University of Denmark,
2800 Kgs. Lyngby, Denmark
}

\date{\today}% It is always \today, today,
             %  but any date may be explicitly specified

% --- Abstract ---
\begin{abstract}
Understanding the spatial distribution of carriers is important for interpreting transport in nanoscale devices. Here, we apply conductance fluctuation tomography to planar selective-area-grown InAs nanowires in both normal-normal and normal-superconductor device geometries. By tracking the evolution of conductance-interference features as a function of magnetic-field strength and orientation, we extract information about the geometry of phase-coherent transport pathways. Using theory to distinguish between bulk-dominated transport, coherent near-surface transport across facets, and transport confined to individual facets. The measurements are consistent with transport dominated by a near-surface accumulation layer in InAs. Devices with normal contacts show behavior consistent with coherent transport across the nanowire apex, whereas hybrid normal-superconductor devices exhibit signatures of more facet-dependent transport. These results demonstrate how universal conductance fluctuations can be used as a tomographic probe of phase-coherent transport pathways in semiconductor nanostructures.
\end{abstract}

\maketitle

% --- Introduction ---
\section*{Introduction} 

By enabling scalable, lithographic definition of nanoscale crystal growth, Selective Area Growth (SAG) has emerged as a powerful technique for the realization of nanowire devices, circuits, and designed crystal lattices~\cite{krizek,Beznasiuk2022}. Prototype devices based on conventional vapour-liquid-solid (VLS) growth of out-of-plane nanowires have, for the past decades, been an important platform for low-temperature quantum device research. SAG enables deterministic in-plane nanowire networks compatible with standard planar processing, providing a potential route towards scalable cryogenic nanowire circuitry~\cite{olvsteins2023cryogenic,Olsteins_reproducibility_2024}. 

The spatial distribution of carriers strongly influences the electrostatic response, scattering environment, and phase-coherent transport properties of nanowire quantum devices. For example, compared to bulk-dominated transport, carriers confined near the nanowire surfaces experience enhanced Rashba spin–orbit coupling due to the breaking of inversion symmetry and strong electric fields at the interfaces which also determines the effective direction of the spin-orbit field, which can be a crucial parameter for functionality driven by spin-orbit coupling \cite{alicea2011non, Nadj_Perge_2010}.  Furthermore, devices dominated by surface transport generally show stronger electrostatic gate coupling, but also increased sensitivity to surface scattering, roughness, disorder, and charged adsorbates, which can reduce mobility and phase coherence. In contrast, bulk-dominated transport is typically associated with higher mobility and more ballistic transport characteristics~\cite{Peng2013SurfaceTransport,vanTilburg2010CoreShell,Wang2013SurfaceRoughness}. InAs is generally expected to host a surface accumulation layer due to surface-state-induced Fermi-level pinning above the conduction-band edge, enhancing the electron density near the nanowire surface~\cite{Olsson}. This surface accumulation is also responsible for the formation of low-resistance ohmic contacts without the need for optimized annealing procedures~\cite{Dayeh,Ford2009}. While surface accumulation in planar InAs systems is well established~\cite{Olsson}, its spatial distribution and effective conductance thickness remain less well understood in confined InAs nanostructures~\cite{degtyarev2017surface} and have not been addressed for in-plane SAG nanowires. Determining the spatial distribution of carriers under realistic cryogenic operating conditions therefore remains an important experimental challenge. 

For coherent quantum devices in a diffusive or quasi-ballistic regime, an interesting option is to utilize the modulation of the interference of the electron wavefunctions in magnetic field for tomographic imaging. Interference loops due to impurity or boundary scattering result in fluctuations in the device conductance, which modulate with the enclosed magnetic flux \cite{gupta_conductance_1994,lee_universal_1985,imry_active_1986, mello_macroscopic_1988,lee_universal_1987,beenakker_random-matrix_1997}. Thus, by tracking interference features as the external field strength $|B|$ and the field orientation is varied, it is possible to infer the geometry of the interference loop and thus the spatial distribution of the conducting system. This method was first demonstrated in metallic nanowires \cite{Scheer1997}, where the anisotropy of universal conductance fluctuations (UCF), was used to distinguish two-dimensional surface-dominated transport from three-dimensional bulk contributions and to demonstrate and study the formation of a surface accumulation layer in VLS nanowires \cite{Liang,Haas2016,Tsand}.

%% Figure 1

\begin{figure*}[t!]
    \centering
    \includegraphics[width=1\textwidth]{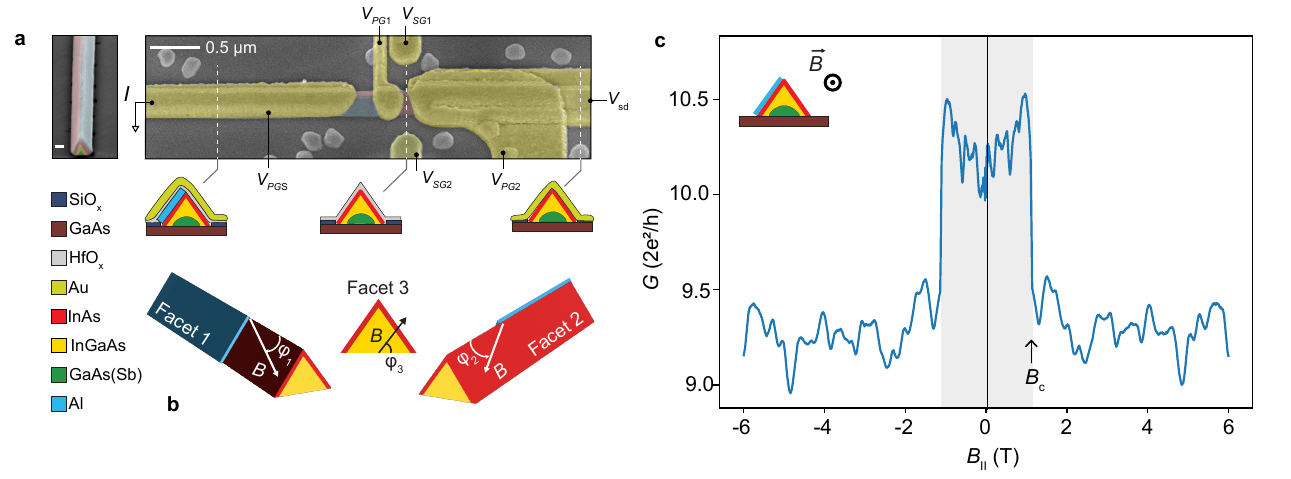}
    \caption{\textbf{a}, False-colored scanning electron micrograph of a typical as-grown nanowire (leftmost inset; scale bar is $100 \, \mathrm{nm}$) and a hybrid normal-superconductor device with schematic cross-sections shown at three positions (rightmost inset). \textbf{b}, Definition of facet planes and angles $\varphi_1, \varphi_2, \varphi_3$ of the magnetic field, $B$, with respect to the nanowire geometry. \textbf{c}, Conductance $G$ as a function of magnetic field applied parallel to the nanowire, $B_\parallel$. $B_c$ indicates the critical field of the superconducting Al.}
    \label{fig:fig1}
\end{figure*}

In VLS nanowires, a surface accumulation layer is expected to give rise to Aharonov-Bohm-type oscillations in a magnetic field applied parallel to the nanowire axis. While such oscillations were experimentally observed in core-shell nanowires~\cite{Haas2016}, they were absent in pure InAs nanowires~\cite{Tsand}, potentially indicating strong scattering or bulk transport. Self-consistent Schrödinger-Poisson simulations have shown that heterostructure design can strongly modify carrier confinement and, in some cases, promote transport channels localized near nanowire surfaces~\cite{daria}. In practice, growth-related intermixing and the incorporation of additional elements may alter this idealized picture and modify the resulting carrier distribution~\cite{Trevisan_2025}. Furthermore, regions where neighboring facets meet may exhibit non-ideal morphology or enhanced defect formation, potentially influencing coherent coupling and transport between adjacent facets~\cite{Espineira}. 

Here we apply UCF tomography to InAs SAG nanowires grown with a triangular cross section on GaAs. The aim is to investigate bulk versus surface transport and the role of coherent coupling across neighboring surface facets. We distinguish between three qualitative transport scenarios: i) transport with significant bulk contribution, ii) near-surface transport with coherent coupling across the apex, and iii) transport along two largely independent surface facets. By comparing normal-normal (NN) and normal-superconducting (NS) devices, we further investigate how a single superconducting facet modifies the underlying phase-coherent transport pathways.

\section*{Methods}

Planar InAs/InGaAs SAG nanowires were grown by selective area molecular beam epitaxy (MBE) based on templates prepared by first covering a GaAs (001) substrate with a $\sim 10\, \mathrm{nm}$ thick SiO$_2$ mask layer, which was then locally etched to expose the GaAs surface in $0.25 \, \times 10 \, \mu \mathrm m $ regions, oriented along the [011] in-plane directions. The growth sequence includes a GaAs(Sb) buffer to prepare the surface, an InGaAs step to accommodate the strain between the GaAs and the subsequent InAs, which has a thickness of 30 nm. Finally, $\sim \, 20\mathrm{nm}$ of epitaxial Al was deposited \emph{in situ} perpendicular to the nanowire axis (at a shallow angle to the GaAs plane), thus covering one side facet of the resulting nanowire.  The details on growth and structural analysis have been reported in Ref. \cite{Beznasiuk2022}.  The nanowires attain triangular cross-sections with $70^\circ$ angle at the apex, and the leftmost inset in Fig.\ 1a  shows a scanning electron micrograph of a typical as-grown nanowire.

In order to perform UCF tomography, electrical devices were fabricated directly on the growth substrate. Using e-beam lithography, the Al half-shell was locally etched, and Ti/Au source and drain contacts were defined to realize devices in two geometries: NN (Ti/Au-nanowire-Ti/Au) and NS (Ti/Au-nanowire-Al). All devices have a channel length of $L\sim250 \, \mathrm{nm}$. The devices were conformally covered by $15 \, \mathrm{nm}$ of HfO$_2$ dielectric layer using atomic layer deposition, followed by fabrication of electrostatic Ti/Au top gates. The rightmost inset in Fig.~\ref{fig:fig1}a shows a fully fabricated NS device and the measurement configuration. The cross sections $1-3$ illustrate the structure at different positions along the device. The device features five top gates, and unless stated otherwise, gates $V_{\mathrm{PG1}}$ and $V_{\mathrm{PG2}}$ are fixed at 1.75~V, where the semiconductor density is far from pinch-off. The remaining gates are grounded.

\begin{figure*}[t]
    \centering
    \includegraphics[width=1\textwidth, page=1]{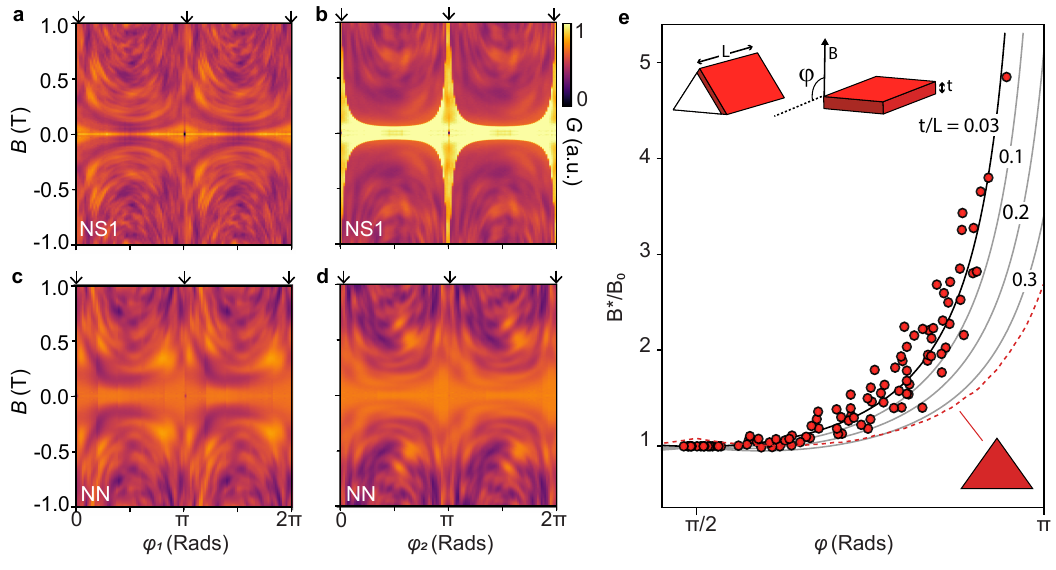}
   \caption{ Differential conductance for Device NS (\textbf{a-b}) and NN (\textbf{c-d}) as a function of $|B|$ and the field angles, $\varphi_1,\varphi_2$. The arrows indicate the orientation of the nanowire axis. \textbf{e}, Magnetic-field positions $B^*(\varphi)$ of interference features extracted from Fig.~\ref{fig:fig2}a-d and normalized by their corresponding value $B_0$ at $\varphi=\pi/2$. Dashed red line shows the expected behavior for a solid triangular nanowire, obtained from numerical simulations. Solid lines show the expected behavior for rectangular box with varying surface accumulation layer thickness-to-length ratio, $t/L$.}  
    \label{fig:fig2}
\end{figure*}

\section*{Results $\&$ Discussion}

Measurements were performed in a dilution cryostat with base temperature, $T \sim$ 20 mK, and equipped with a superconducting vector magnet. We denote by Facet 1 and Facet 2 the two planes of the side facets, and by F3 the cross-sectional plane orthogonal to both F1 and F2. Figure \ref{fig:fig1}b defines the angles $\varphi_1,\varphi_2,\varphi_3$ used to describe the angles of the field, $B$, when rotating within F1,F2, and F3, respectively. In addition, $B_\parallel$ denotes a field applied parallel to the nanowire axis. The two-terminal conductance $G$ was measured using standard lock-in techniques with an \emph{ac} excitation of $10 \, \mu \mathrm V$. The cryostat line resistance and the smaller contact resistance were subtracted from the measured conductance (see Supplemental Material section I). In total, three devices were measured at low temperatures, which are denoted NS1, NS2, and NN.

Figure~\ref{fig:fig1}c shows $G(B_\parallel)$ for the device NS1. Reproducible, $B$-symmetric, aperiodic fluctuations are observed with a magnitude of $\sim e^2/h$, which are characteristic of UCF \cite{lee_universal_1985, gupta_conductance_1994,beenakker_random-matrix_1993}. The decrease in conductance at $B_\parallel \sim 1.2 \, \mathrm T$ corresponds to the critical field $B^c_\parallel$ of the Al contact.

Figure \ref{fig:fig2}a-d shows measurements of $G(B)$ for different orientations of the field and for two different devices NS1 and NN (See Supplemental section II for additional data). Panels a,b (c,d) show the result for devices NS (NN) when the field was rotated in the plane of F1 (a,c) and F2 (b,d) where  $\varphi_1$ and $\varphi_2$ denote the angle of $B$ with respect to the nanowire axis (cf. Fig.\ 1b). Since for the two-terminal devices, $G(B)=G(-B)$, as confirmed in Fig. \ref{fig:fig1}c, the measurements were performed for $B>0$ and $\varphi \in [0,\pi]$ and repeated to the range $-1 \, \mathrm T \le B \le 1 \, \mathrm T$ for better visualization. The conductance fluctuations depend on the field orientations, and form a pattern where features can be followed between different $\varphi$ with a tendency to shift towards higher fields around specific angles. In Fig.~\ref{fig:fig2}a, $B$ remains in-plane of the superconducting aluminum film for all $\varphi_{\mathrm{1}}$, and the NS device stays in the superconducting state. In Figs.~\ref{fig:fig2}b, however, the field is parallel to the Al-film for $\varphi_2 = 0, \pi, 2\pi$ and acquires an increasing perpendicular component for $\varphi_2 \rightarrow \pi/2, 3\pi/2$, which suppresses superconductivity in the Al contact above a critical field. As a result, in addition to the conductance fluctuations, a low-field region of enhanced conductance appears, corresponding to the superconducting state of the NS device (cf.\ Fig. \ref{fig:fig1}c).

The patterns in Fig.~\ref{fig:fig2}a-d are characteristic of angle-dependent conductance fluctuations and can be qualitatively understood following Refs.~\cite{Tsand,Haas2016} by associating conductance features with interference of quasi-two-dimensional loops enclosing a magnetic flux. For a given interference loop, the same feature appears at different magnitudes of the magnetic fields for different field orientations, such that a constant enclosed flux is maintained. This leads to an angular dependence of the form $B(\varphi) \propto {\cos\!\left(\varphi-\varphi_0\right)^{-1}}$, where $\varphi - \varphi_0$ is the angle between the magnetic field and the normal to the loop plane. Consequently, interference features shift toward higher fields as the orientation of $B$ approaches the plane of the loops.

In the case where the SAG nanowire is best described as (i) either having bulk transport or (ii) having two planar near-surface conducting facets that meet at the apex, rotating $B$ within the facet planes F1(F2) (angles $\varphi_1,\varphi_2$) thus modulates interference of transport projected on facets F2(F1) with features shifting towards higher $|B|$ when the field is tilted towards the nanowire axis ($\varphi_{1,2} = 0,\pi$). An interference feature appearing at $B_0$ for $\varphi = \pi/2$ will shift towards higher fields $B^*(\varphi) = B_0/A_p(\varphi)$ as the field is tilted. Here, $A_p$ is the projected loop area. This is qualitatively consistent with the results in Fig.\ \ref{fig:fig2}a-d where the arrows indicate the orientation of the nanowire axis. In case (i) the relevant normalized projected area, $A_p^{(i)}(\varphi_1)$, is that of a solid triangular nanowire, which was obtained by simulations and $B_0/A_p^{(i)}(\varphi_1)$ is shown by the dashed line in Fig.\ \ref{fig:fig2}e (see the supplementary information section III). In scenario (ii) the facet is modeled as a rectangular box with a length $L = 250\,\mathrm{nm}$, width $W = 200\,\mathrm{nm}$, and thickness $t$ corresponding to a finite effective thickness of the surface accumulation layer. In this case, the projected area for a direction forming an angle $\varphi$ to the nanowire axis is then $A^{(ii)}_\perp(\varphi) = WL\times A^{(ii)}_p(\varphi)$ where $A_p^{(ii)}(\varphi) \propto \sin(\varphi) + t/L\cos(\varphi)$. Figure \ref{fig:fig2}e shows the field values $B^*(\varphi)$ normalized by $B_0$ of all interference features which, in Fig.\ \ref{fig:fig2}a-d, can be tracked from $\varphi = \pi/2$ (see details in SI Section 4). The steep increase of $B^*(\varphi)$ for $\varphi \rightarrow \pi$ is inconsistent with $1/A^{(i)}_p(\varphi_1)$ (red dashed line) and therefore rules out scenario (i) of a nanowire with significant bulk transport. Solid lines in Fig.\ \ref{fig:fig2}e show $1/A^{(ii)}_p(\varphi_1)$ for different values of the accumulation layer thickness, $t/L$. Fitting to the data yields $t/L = (0.02 \pm 0.03)$ corresponding to a thickness of the accumulation layer below $13 \,\mathrm{nm}$. This value is consistent with previous estimations of the InAs surface-channel thickness \cite{daria}. 

Having established that transport is confined to the nanowire facets, we now consider the two limiting scenarios (ii) and (iii) discussed above. For rotations in the cross-sectional plane F3 (angle $\varphi_3$), the field modulates transport on both facets, and the two scenarios are expected to produce qualitatively different interference patterns. Numerical simulations of the corresponding projected loop areas, shown in Supplemental Section III and Fig.~S4b, illustrate these distinct behaviors. If full coherent transport across the apex occurs (scenario ii), there is, on average, no preference for either facet, and no clearly separated diverging families of interference features are expected in maps of $G(|B|,\varphi_3)$. In contrast, if coherent transport between the facets is limited (scenario iii), two diverging families should appear around $\varphi_3=\pm35^\circ$, corresponding to the $70^\circ$ angle between the two facets (cf.\ Fig.~1b).
\begin{figure}[t]
    \centering
    \includegraphics[width=0.7\columnwidth, page=2]{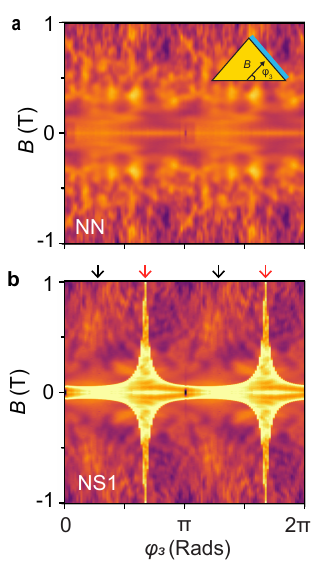}
    \caption{Differential conductance for Device NN (\textbf{a}) and NS1 (\textbf{b}) as a function of $|B|$ and the field angle, $\varphi_3$, within the plane normal to the nanowire axis. Arrow in panel (b) indicate the orientatins of the two facet planes forming an angle of $\sim 70^\circ$}
    \label{fig:fig3}
\end{figure}

Figure~\ref{fig:fig3}a shows the measurement of $G(|B|,\varphi_3)$ for the NN device. No clear diverging families are observed, suggesting coherent transport across the apex. This is consistent with previous observations of phase-coherent inter-facet transport in InAs core-shell VLS nanowires \cite{Haas2016}. Interestingly, the corresponding measurements of NS1 in Fig.~\ref{fig:fig3}b exhibit interference features with a tendency to diverge around $\varphi_3=\pm35^\circ$, as indicated by the red and black arrows. Although the origin of this difference between NS and NN devices remains unclear, it suggests that transport in single-facet superconducting devices may differ qualitatively from that in devices contacted by normal evaporated metal leads.

\section*{Conclusion}

In conclusion, we have investigated low-temperature conductance fluctuations in planar selective-area-grown InAs nanowires with both NN and NS device geometries. By analyzing the magnetic field strength and angular dependence of the fluctuations, characteristic signatures of quasi-two-dimensional interference loops were identified, enabling extraction of information about the spatial distribution of carriers within the nanowire cross-section. The results are consistent with significant bulk transport and instead support transport dominated by a near-surface accumulation layer with an estimated thickness below 13 nm, consistent with expectations for surface accumulation in InAs. Furthermore, the NN device shows evidence of coherent transport between the two side facets of the nanowire, whereas the NS device suggests reduced coherent transport across the apex. These results demonstrate how UCF tomography can be used to probe the spatial distribution of phase-coherent transport pathways in semiconductor nanostructures. The work further highlights the sensitivity of coherent transport to superconducting hybridization and facet-dependent confinement in selectively grown InAs nanowires, a materials platform currently being explored for scalable nanowire quantum devices and quantum circuitry \cite{krizek,olvsteins2023cryogenic,Olsteins_reproducibility_2024}.

%near-surface transport with coherent scattering across facets. Furthermore, the absence of clearly resolved crossing divergence families in the normal-state data disfavors transport dominated by two fully independent facets. A quantitative analysis based on a finite-thickness slab model yields an effective conducting thickness of $t \sim 8$-$25\,\mathrm{nm}$, in good agreement with the expected InAs surface-channel thickness. This is consistent with phase-coherent transport being confined to a near-surface region of the nanowire. In hybrid NS devices, we observe indications of a weak facet asymmetry, suggesting that single-facet superconducting contacts can influence the distribution of phase-coherent trajectories. Our results demonstrate that angle-resolved UCF provides a simple tool to extract geometrical information about phase-coherent transport pathways.

\bibliographystyle{apsrev4-2}
\bibliography{CPBib}

@PREAMBLE{"\providecommand{\noopsort}[1]{} \providecommand{\singleletter}[1]{#1}%"}

@article{Trevisan_2025, 
    title={Iso-electronic Sb impurities in GaAs studied by cross-sectional scanning tunneling microscopy},
    volume={137},
    ISSN={0021-8979},
    DOI={10.1063/5.0263956}, 
    number={19},
    journal={Journal of Applied Physics}, 
    author={Trevisan, Aurelia and Hodgson, Peter D. and Alvarado-César, Francisco and Hayne, Manus and Beanland, Richard and Koenraad, Paul M.}, 
    year={2025},
    pages={195702}
}

@article{olvsteins2023cryogenic,
  title={Cryogenic multiplexing using selective area grown nanowires},
  author={Ol{\v{s}}teins, D{\=a}gs and Nagda, Gunjan and Carrad, Damon J and Beznasyuk, Daria V and Petersen, Christian EN and Mart{\'\i}-S{\'a}nchez, Sara and Arbiol, Jordi and Jespersen, Thomas S},
  journal={Nature communications},
  volume={14},
  number={1},
  pages={7738},
  year={2023},
  publisher={Nature Publishing Group UK London}
}

@article{alicea2011non,
  title={Non-Abelian statistics and topological quantum information processing in 1D wire networks},
  author={Alicea, Jason and Oreg, Yuval and Refael, Gil and Von Oppen, Felix and Fisher, Matthew PA},
  journal={Nature Physics},
  volume={7},
  number={5},
  pages={412--417},
  year={2011},
  publisher={Nature Publishing Group}
}

@article{Olsteins_reproducibility_2024,
 title={Statistical Reproducibility of Selective Area Grown InAs Nanowire Devices}, 
 volume={24},
 ISSN={1530-6984},
 DOI={10.1021/acs.nanolett.4c01038},
 number={22},
 journal={Nano Letters},
 author={Olsteins, Dags and Nagda, Gunjan and Carrad, Damon J. and Beznasyuk, Daria V. and Petersen, Christian E. N. and Martí-Sánchez, Sara and Arbiol, Jordi and Jespersen, Thomas Sand}, 
 year={2024},
 pages={6553–6559}
}

@article{Peng2013SurfaceTransport,
  author  = {Peng, Weina and Aksamija, Zlatan and Scott, Shelley A.
             and Endres, James J. and Savage, Donald E.
             and Knezevic, Irena and Eriksson, Mark A.
             and Lagally, Max G.},
  title   = {Probing the Electronic Structure at Semiconductor Surfaces Using Charge Transport in Nanomembranes},
  journal = {Nature Communications},
  volume   = {4},
  pages    = {1339},
  year     = {2013},
  doi      = {10.1038/ncomms2350}
}

@article{Dayeh,
author = {Dayeh, Shadi A. and Aplin, David P. R. and Zhou, Xiaotian and Yu, Paul K. L. and Yu, Edward T. and Wang, Deli},
title = {High Electron Mobility InAs Nanowire Field-Effect Transistors},
journal = {Small},
volume = {3},
number = {2},
pages = {326-332},
doi = {https://doi.org/10.1002/smll.200600379},
year = {2007}
}

@article{vanTilburg2010CoreShell,
  author  = {van Tilburg, J. W. W. and Algra, R. E.
             and Immink, G. and Verheijen, M. A.
             and Bakkers, E. P. A. M. and Kouwenhoven, L. P.},
  title   = {Surface-Passivated InAs/InP Core--Shell Nanowires},
  journal = {Semiconductor Science and Technology},
  volume   = {25},
  number   = {2},
  pages    = {024011},
  year     = {2010},
  doi      = {10.1088/0268-1242/25/2/024011}
}

@article{Wang2013SurfaceRoughness,
  author  = {Wang, Jian and Gudiksen, Mark S.},
  title   = {Surface Roughness Induced Electron Mobility Degradation in InAs Nanowires},
  journal = {Nanotechnology},
  volume   = {24},
  number   = {37},
  pages    = {375202},
  year     = {2013},
  doi      = {10.1088/0957-4484/24/37/375202}
}

@article{degtyarev2017surface,
  author  = {Degtyarev, V. E. and Khazanova, S. V. and Demarina, N. V.},
  title   = {Features of electron gas in InAs nanowires imposed by interplay between nanowire geometry, doping and surface states},
  journal = {Scientific Reports},
  volume  = {7},
  pages   = {3411},
  year    = {2017},
  doi     = {10.1038/s41598-017-03415-3}
}

@article{Olsson,
  title = {Charge Accumulation at InAs Surfaces},
  author = {Olsson, L. \"O. and Andersson, C. B. M. and H\aa{}kansson, M. C. and Kanski, J. and Ilver, L. and Karlsson, U. O.},
  journal = {Phys. Rev. Lett.},
  volume = {76},
  issue = {19},
  pages = {3626--3629},
  numpages = {0},
  year = {1996},
  month = {May},
  publisher = {American Physical Society},
  doi = {10.1103/PhysRevLett.76.3626},
  url = {https://link.aps.org/doi/10.1103/PhysRevLett.76.3626}
}

@article{Espineira,
  title = {Selective area growth rates of III-V nanowires},
  author = {Cachaza, Martin Espi\~neira and Christensen, Anna Wulff and Beznasyuk, Daria and S\ae{}rkj\ae{}r, Tobias and Madsen, Morten Hannibal and Tanta, Rawa and Nagda, Gunjan and Schuwalow, Sergej and Krogstrup, Peter},
  journal = {Phys. Rev. Mater.},
  volume = {5},
  issue = {9},
  pages = {094601},
  numpages = {7},
  year = {2021},
  month = {Sep},
  publisher = {American Physical Society},
  doi = {10.1103/PhysRevMaterials.5.094601},
  url = {https://link.aps.org/doi/10.1103/PhysRevMaterials.5.094601}
}

@article{Tsand,
  title = {Probing the spatial electron distribution in InAs nanowires by anisotropic magnetoconductance fluctuations},
  author = {Jespersen, T. S. and Hauptmann, J. R. and S\o{}rensen, C. B. and Nyg\aa{}rd, J.},
  journal = {Phys. Rev. B},
  volume = {91},
  issue = {4},
  pages = {041302},
  numpages = {5},
  year = {2015},
  month = {Jan},
  publisher = {American Physical Society},
  doi = {10.1103/PhysRevB.91.041302},
  url = {https://link.aps.org/doi/10.1103/PhysRevB.91.041302}
}

@article{Liang,
  title = {Anisotropic magnetoconductance of a InAs nanowire: Angle-dependent suppression of one-dimensional weak localization},
  author = {Liang, Dong and Du, Juan and Gao, Xuan P. A.},
  journal = {Phys. Rev. B},
  volume = {81},
  issue = {15},
  pages = {153304},
  numpages = {4},
  year = {2010},
  month = {Apr},
  publisher = {American Physical Society},
  doi = {10.1103/PhysRevB.81.153304},
  url = {https://link.aps.org/doi/10.1103/PhysRevB.81.153304}
}

@article{Haas2016,
  author  = {Haas, Fabian and Wenz, Tobias and Zellekens, Patrick and Demarina, Nataliya and Rieger, Torsten and Lepsa, Mihail and Grützmacher, Detlev and Lüth, Hans and Schäpers, Thomas},
  title   = {Angle‐dependent magnetotransport in {GaAs}/{InAs} core/shell nanowires},
  journal = {Scientific Reports},
  volume  = {6},
  pages   = {24573},
  year    = {2016},
  doi     = {10.1038/srep24573}
}

@article{Scheer1997,
  author  = {Scheer, E. and von Löhneysen, H. and Mirlin, A. D. and Wölfle, P. and Hein, H.},
  title   = {Angular Dependence of Universal Conductance Fluctuations in Noble‐Metal Nanowires},
  journal = {Physical Review Letters},
  volume  = {78},
  number  = {17},
  pages   = {3362--3365},
  year    = {1997},
  doi     = {10.1103/PhysRevLett.78.3362}
}

@article{Beznasiuk2022,
  title = {Doubling the mobility of InAs/InGaAs selective area grown nanowires},
  author = {Beznasyuk, Daria V. and Mart\'{\i}-S\'anchez, Sara and Kang, Jung-Hyun and Tanta, Rawa and Rajpalke, Mohana and Stankevi\ifmmode \check{c}\else \v{c}\fi{}, Toma\ifmmode \check{s}\else \v{s}\fi{} and Christensen, Anna Wulff and Spadaro, Maria Chiara and Bergamaschini, Roberto and Maka, Nikhil N. and Petersen, Christian Emanuel N. and Carrad, Damon J. and Jespersen, Thomas Sand and Arbiol, Jordi and Krogstrup, Peter},
  journal = {Phys. Rev. Mater.},
  volume = {6},
  issue = {3},
  pages = {034602},
  numpages = {9},
  year = {2022},
  month = {Mar},
  publisher = {American Physical Society},
  doi = {10.1103/PhysRevMaterials.6.034602},
  url = {https://link.aps.org/doi/10.1103/PhysRevMaterials.6.034602}
}

@article{Ford2009,
  author = {Ford, A. C. and Ho, J. C. and Chueh, Y.-L. and Tseng, Y.-C. and Fan, Z. and Guo, J. and Bokor, J. and Javey, A.},
  title = {Diameter-Dependent Electron Mobility of InAs Nanowires},
  journal = {Nano Letters},
  volume = {9},
  number = {1},
  pages = {360--365},
  year = {2009},
  doi = {10.1021/nl803154m}
}

@article{beenakker_random-matrix_1997,
	title = {Random-matrix theory of quantum transport},
	volume = {69},
	url = {https://link.aps.org/doi/10.1103/RevModPhys.69.731},
	doi = {10.1103/RevModPhys.69.731},
	number = {3},
	urldate = {2024-05-22},
	journal = {Reviews of Modern Physics},
	author = {Beenakker, C. W. J.},
	month = jul,
	year = {1997},
	note = {Publisher: American Physical Society},
	pages = {731--808},
}

@article{beenakker_random-matrix_1993,
	title = {Random-matrix theory of mesoscopic fluctuations in conductors and superconductors},
	volume = {47},
	url = {https://link.aps.org/doi/10.1103/PhysRevB.47.15763},
	doi = {10.1103/PhysRevB.47.15763},
	number = {23},
	urldate = {2024-05-22},
	journal = {Physical Review B},
	author = {Beenakker, C. W. J.},
	month = jun,
	year = {1993},
	note = {Publisher: American Physical Society},
	pages = {15763--15775},
}

@article{krizek,
  title = {Field effect enhancement in buffered quantum nanowire networks},
  author = {Krizek, Filip and Sestoft, Joachim E. and Aseev, Pavel and Marti-Sanchez, Sara and Vaitiek\ifmmode \dot{e}\else \.{e}\fi{}nas, Saulius and Casparis, Lucas and Khan, Sabbir A. and Liu, Yu and Stankevi\ifmmode \check{c}\else \v{c}\fi{}, Toma\ifmmode \check{s}\else \v{s}\fi{} and Whiticar, Alexander M. and Fursina, Alexandra and Boekhout, Frenk and Koops, Rene and Uccelli, Emanuele and Kouwenhoven, Leo P. and Marcus, Charles M. and Arbiol, Jordi and Krogstrup, Peter},
  journal = {Phys. Rev. Mater.},
  volume = {2},
  issue = {9},
  pages = {093401},
  numpages = {8},
  year = {2018},
  month = {Sep},
  publisher = {American Physical Society},
  doi = {10.1103/PhysRevMaterials.2.093401},
  url = {https://link.aps.org/doi/10.1103/PhysRevMaterials.2.093401}
}

@article{daria,
  title = {Doubling the mobility of InAs/InGaAs selective area grown nanowires},
  author = {Beznasyuk, Daria V. and Mart\'{\i}-S\'anchez, Sara and Kang, Jung-Hyun and Tanta, Rawa and Rajpalke, Mohana and Stankevi\ifmmode \check{c}\else \v{c}\fi{}, Toma\ifmmode \check{s}\else \v{s}\fi{} and Christensen, Anna Wulff and Spadaro, Maria Chiara and Bergamaschini, Roberto and Maka, Nikhil N. and Petersen, Christian Emanuel N. and Carrad, Damon J. and Jespersen, Thomas Sand and Arbiol, Jordi and Krogstrup, Peter},
  journal = {Phys. Rev. Mater.},
  volume = {6},
  issue = {3},
  pages = {034602},
  numpages = {9},
  year = {2022},
  month = {Mar},
  publisher = {American Physical Society},
  doi = {10.1103/PhysRevMaterials.6.034602},
  url = {https://link.aps.org/doi/10.1103/PhysRevMaterials.6.034602}
}

@article{Nadj_Perge_2010,
   title={Spin–orbit qubit in a semiconductor nanowire},
   volume={468},
   ISSN={1476-4687},
   url={http://dx.doi.org/10.1038/nature09682},
   DOI={10.1038/nature09682},
   number={7327},
   journal={Nature},
   publisher={Springer Science and Business Media LLC},
   author={Nadj-Perge, S. and Frolov, S. M. and Bakkers, E. P. A. M. and Kouwenhoven, L. P.},
   year={2010},
   month=dec, pages={1084–1087} }

@article{lee_universal_1985,
	title = {Universal {Conductance} {Fluctuations} in {Metals}},
	volume = {55},
	url = {https://link.aps.org/doi/10.1103/PhysRevLett.55.1622},
	doi = {10.1103/PhysRevLett.55.1622},
	number = {15},
	urldate = {2024-05-22},
	journal = {Physical Review Letters},
	author = {Lee, P. A. and Stone, A. Douglas},
	month = oct,
	year = {1985},
	note = {Publisher: American Physical Society},
	pages = {1622--1625},
}

@article{lee_universal_1987,
	title = {Universal conductance fluctuations in metals: {Effects} of finite temperature, interactions, and magnetic field},
	volume = {35},
	shorttitle = {Universal conductance fluctuations in metals},
	url = {https://link.aps.org/doi/10.1103/PhysRevB.35.1039},
	doi = {10.1103/PhysRevB.35.1039},
	number = {3},
	urldate = {2024-05-22},
	journal = {Physical Review B},
	author = {Lee, P. A. and Stone, A. Douglas and Fukuyama, H.},
	month = jan,
	year = {1987},
	note = {Publisher: American Physical Society},
	pages = {1039--1070},
}

@article{imry_active_1986,
	title = {Active {Transmission} {Channels} and {Universal} {Conductance} {Fluctuations}},
	volume = {1},
	issn = {0295-5075},
	url = {https://dx.doi.org/10.1209/0295-5075/1/5/008},
	doi = {10.1209/0295-5075/1/5/008},
	
	number = {5},
	urldate = {2024-05-22},
	journal = {Europhysics Letters},
	author = {Imry, Y.},
	month = mar,
	year = {1986},
	pages = {249},
}

@article{mello_macroscopic_1988,
	title = {Macroscopic approach to universal conductance fluctuations in disordered metals},
	volume = {60},
	url = {https://link.aps.org/doi/10.1103/PhysRevLett.60.1089},
	doi = {10.1103/PhysRevLett.60.1089},
	number = {11},
	urldate = {2024-05-22},
	journal = {Physical Review Letters},
	author = {Mello, Pier A.},
	month = mar,
	year = {1988},
	note = {Publisher: American Physical Society},
	pages = {1089--1092},
}

@article{gupta_conductance_1994,
	title = {Conductance fluctuations in mesoscopic conductors at low temperatures},
	volume = {41},
	copyright = {https://ieeexplore.ieee.org/Xplorehelp/downloads/license-information/IEEE.html},
	issn = {00189383},
	url = {http://ieeexplore.ieee.org/document/333828/},
	doi = {10.1109/16.333828},

	number = {11},
	urldate = {2024-05-31},
	journal = {IEEE Transactions on Electron Devices},
	author = {Gupta, M.S.},
	month = nov,
	year = {1994},
	pages = {2093--2106},
}

\end{document}

\title{Supplemental Material for:\\
\textit{Tomography of Transport Pathways in Selective-Area-Grown Nanowires Using Angle-Resolved Conductance Fluctuations}}

\author{Christian E. N. Petersen}
\email{cenpe@dtu.dk}

\affiliation{
Department of Energy Conversion and Storage,
Technical University of Denmark,
2800 Kgs. Lyngby, Denmark
}

\author{Damon J. Carrad}
\affiliation{
Department of Energy Conversion and Storage,
Technical University of Denmark,
2800 Kgs. Lyngby, Denmark
}
\author{Daria Beznasyuk}
\affiliation{
Department of Energy Conversion and Storage,
Technical University of Denmark,
2800 Kgs. Lyngby, Denmark
}
\author{Thomas S. Jespersen}
\affiliation{
Department of Energy Conversion and Storage,
Technical University of Denmark,
2800 Kgs. Lyngby, Denmark
}

\date{\today}

\maketitle

% ===========================================================
\section{Subtracting the Series Resistance}

\begin{figure}[t]
    \centering
    \includegraphics[width=0.95\linewidth]{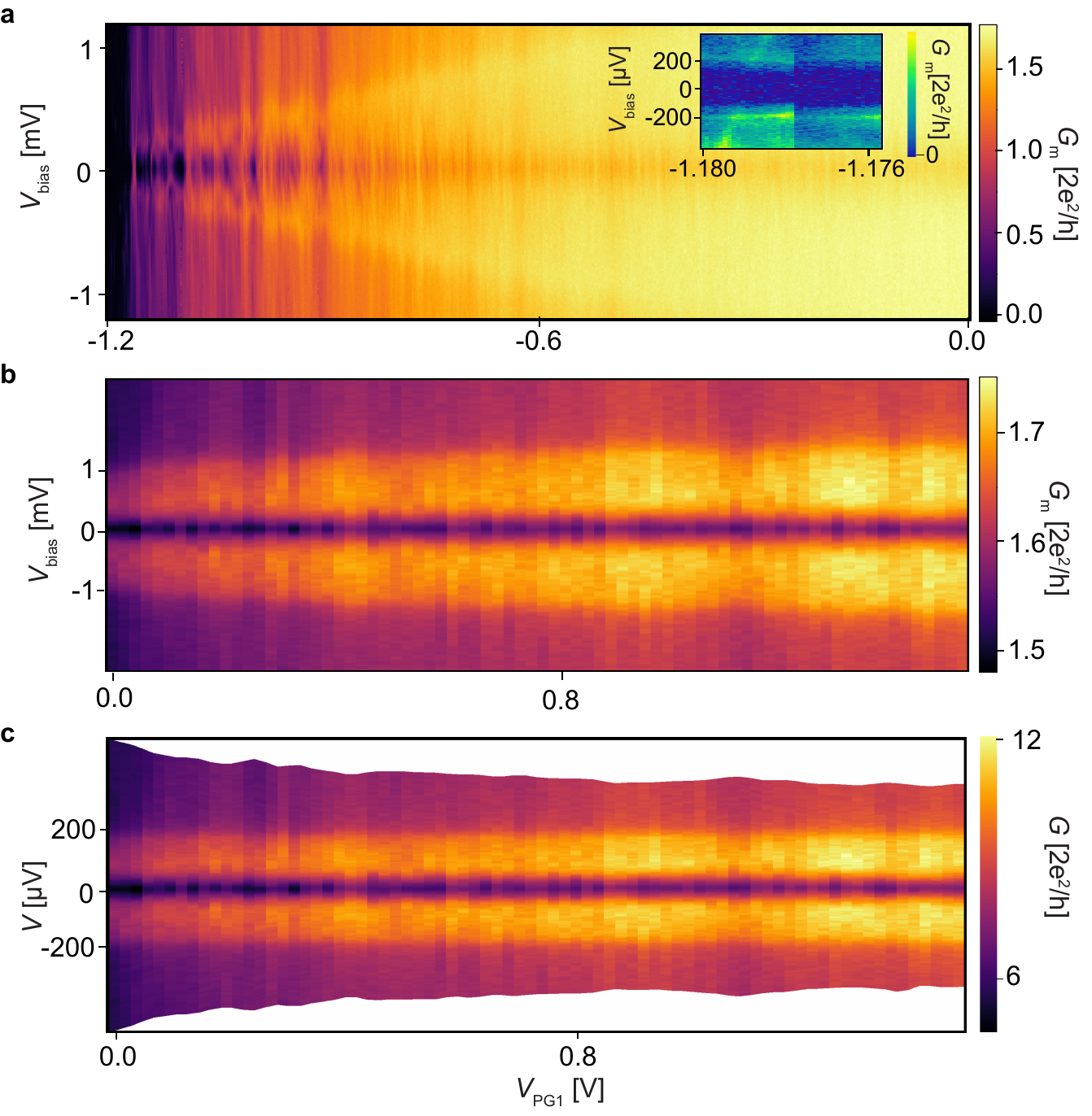}
    \caption{
    \textbf{Series-resistance correction.}
    \textbf{a}, Bias spectroscopy near the pinch-off region, with the inset showing $\Delta/e \sim 200~\mu\mathrm{V}$. 
    $G_\mathrm{m}$ is the measured conductance before subtracting $R_s$.
    \textbf{b}, Bias spectroscopy at higher gate voltages, where the apparent broadening of $\Delta/e$ arises from the voltage drop across a finite $R_s$.  
    \textbf{c}, Bias spectroscopy after redefining the voltage and conductance axes using $R_s \approx 6.35~\mathrm{k}\Omega$, showing a gate-independent value of $\Delta/e$.
    }
    \label{fig:S_series_resistance}
\end{figure}

Figure~\ref{fig:S_series_resistance}a shows bias spectroscopy near the pinch-off region of the semiconductor. 
The inset highlights the superconducting gap divided by the elementary charge, 
$\Delta/e \sim 200~\mu\mathrm{V}$. 
Figure~\ref{fig:S_series_resistance}b presents bias spectroscopy at higher gate voltages, 
where the subgap conductance increases due to stronger coupling between the N and S regions of the nanowire. 
The voltage drop across the device, $V$, and thus its resistance $R$, varies with gate tuning.  

The series resistance $R_s$ can be extracted from the evolution of the gap feature using the voltage-divider relation
\begin{equation}
    V = R \frac{V_\mathrm{bias}}{R_s + R},
    \label{eq:S_voltage_divider}
\end{equation}
or equivalently,
\begin{equation}
    V = V_\mathrm{bias} - R_s I_\mathrm{dc}.
    \label{eq:S_device_voltage}
\end{equation}

At low gate voltages, where $R \gg R_s$, $V \approx V_\mathrm{bias}$, and the superconducting gap appears at 
$\Delta/e \approx 200~\mu\mathrm{V}$. 
The apparent broadening of $\Delta/e$ with increasing gate voltage arises because a smaller device resistance relative to $R_s$ requires a larger applied bias to achieve the same voltage drop across the device.  

We estimate $R_s$ by analyzing the conductance near the edges of the broadened gap at high gate voltage, where the voltage drop across the device remains $V\sim200~\mu\mathrm{V}$. This gives
\begin{equation}
    R_s = \frac{V_\mathrm{bias} - \Delta/e}{I_\mathrm{dc}}.
    \label{eq:S_series_resistance}
\end{equation}

From this analysis, we obtain $R_s \approx 6.35~\mathrm{k}\Omega$. 
Figure~\ref{fig:S_series_resistance}c shows that $\Delta/e$ remains constant as a function of PG1 when Eq.~\eqref{eq:S_device_voltage} is applied, confirming a fixed series resistance.

\clearpage

% ===========================================================
\section{Additional Angle-Resolved Conductance Maps}

\begin{figure}[h!]
    \centering
    \includegraphics[width=0.98\linewidth]{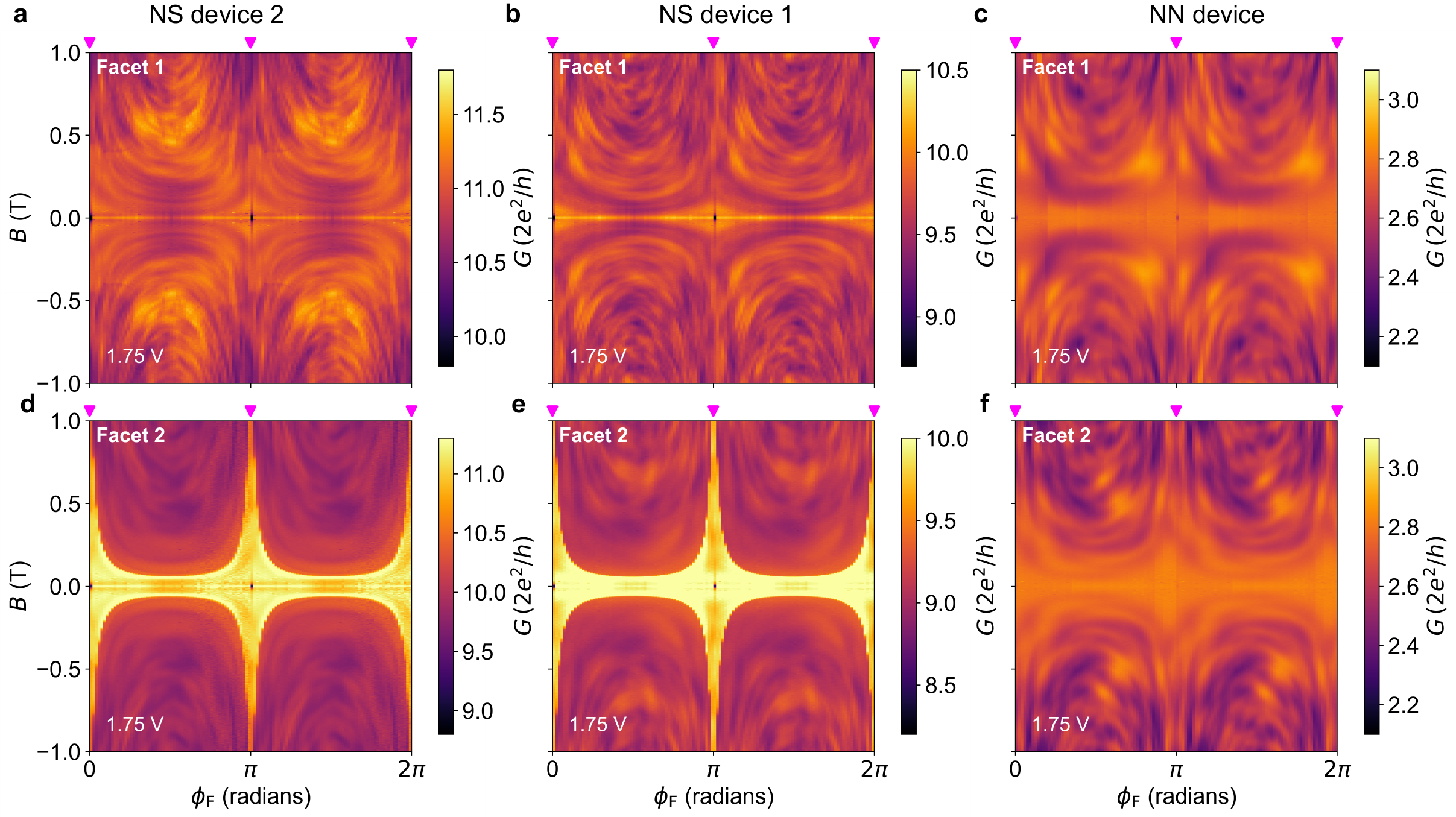}
    \caption{
    \textbf{Angle-resolved conductance maps for rotations in F1 and F2.}
    $G(B,\varphi)$ measured for devices NS1, NS2, and NN as a function of magnetic-field magnitude $B$ and orientation $\varphi_{\mathrm{F}}$ for rotations within the facet planes F1 and F2. The gate configuration was fixed to $V_{\mathrm{PG1}} = V_{\mathrm{PG2}} = 1.75~\mathrm{V}$.
    }
    \label{fig:S_facet_maps}
\end{figure}

\begin{figure}[t]
    \centering
    \includegraphics[width=0.98\linewidth]{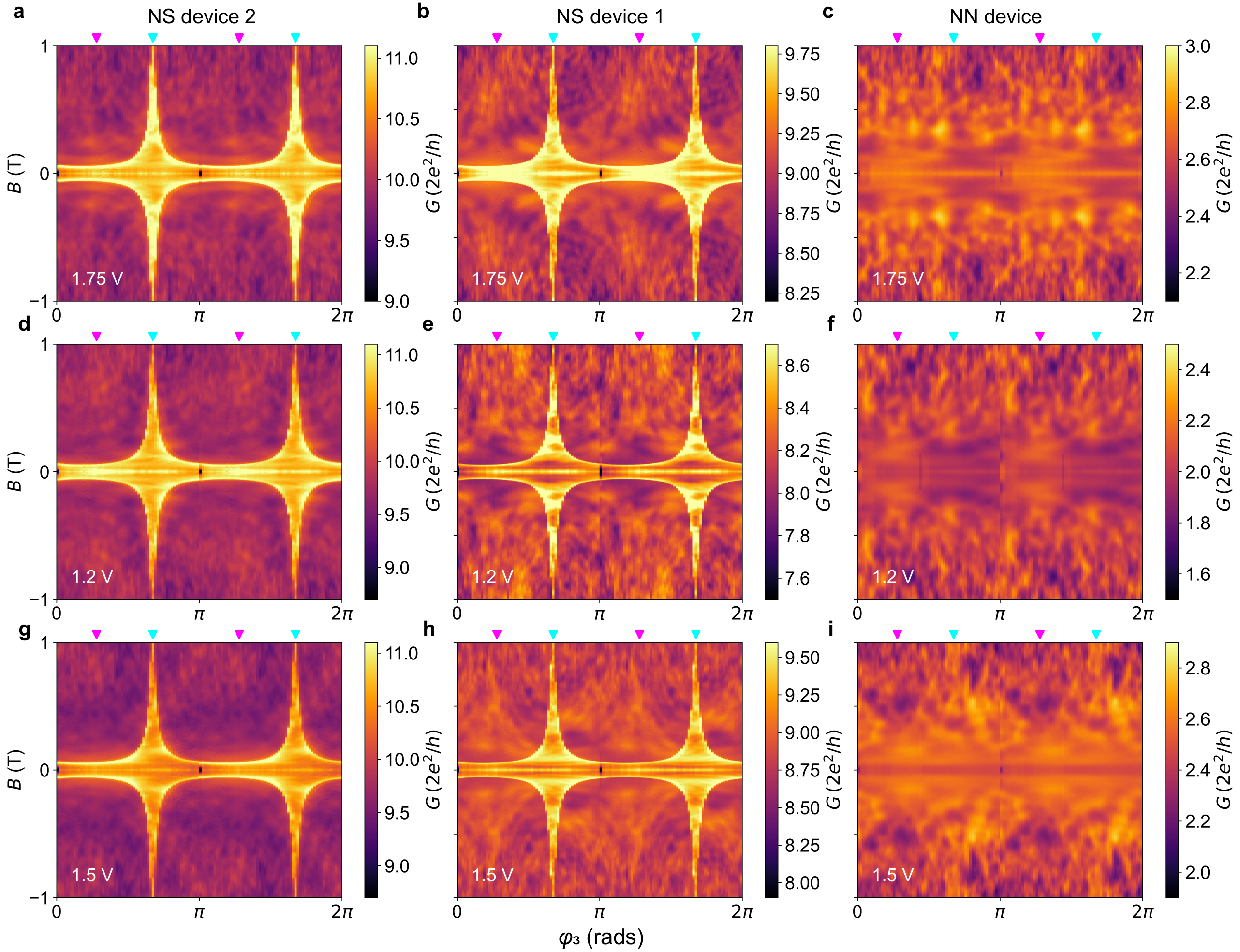}
    \caption{
    \textbf{Angle-resolved conductance maps for rotations in F3.}
    $G(B,\varphi_3)$ measured for rotations in the cross-sectional plane F3 for different gate voltages $V_{\mathrm G}=1.75~\mathrm{V}$, $1.5~\mathrm{V}$, and $1.2~\mathrm{V}$.
    }
    \label{fig:S_tri_maps}
\end{figure}

This section presents the full set of angle-resolved conductance measurements for devices NS1, NS2, and NN obtained for different gate configurations and magnetic-field rotation planes (see Fig. \ref{fig:S_facet_maps} and Fig. \ref{fig:S_tri_maps}). The measurements complement the data shown in the main text and demonstrate the reproducibility of the observed angular conductance fluctuation patterns across devices and gate settings.

Measurements for rotations in F3 were performed at different gate voltages to investigate whether changes in carrier density and electrostatic confinement modify the qualitative character of the interference patterns, and potentially drive a transition between the transport scenarios (i)-(iii) discussed in the main text. Within the accessible gate range, the overall angular dependence remains qualitatively similar, and the measurements do not provide clear evidence for a transition between the different transport regimes. More extreme gate settings or devices with stronger electrostatic tunability may be required to induce more pronounced changes in the transport geometry.

\clearpage

\section{Monte Carlo simulation of inverse projected areas}

To qualitatively compare the experimentally observed angular conductance fluctuation patterns with different transport geometries, we numerically calculated the inverse projected area $1/A_p$ for the three transport scenarios discussed in the main text. The projected area was evaluated using a Monte Carlo ray-casting approach.

For each magnetic-field orientation $\mathbf{B}$, rays parallel to the field direction were sampled through a plane perpendicular to $\mathbf{B}$. The fraction of rays intersecting the model geometry determines the projected area $A_p$. Rotations corresponding to $\varphi_1$ and $\varphi_3$ were simulated by rotating the field direction within the corresponding planes defined in Fig.~1b of the main text.

Three geometries were considered: (i) a solid triangular prism representing homogeneous bulk transport, (ii) two connected finite-thickness facets representing coherent near-surface transport across the apex, and (iii) two independent finite-thickness facets representing transport confined to separate facets without coherent coupling across the apex. For all simulations, a characteristic loop length $L_{\mathrm{MC}}=250~\mathrm{nm}$ and facet width $W_{\mathrm{MC}}=200~\mathrm{nm}$ were used, while the finite-thickness facet models employed a thickness $t_{\mathrm{MC}}=15~\mathrm{nm}$.

\begin{figure}[h!]
    \centering
    \includegraphics[width=1\linewidth]{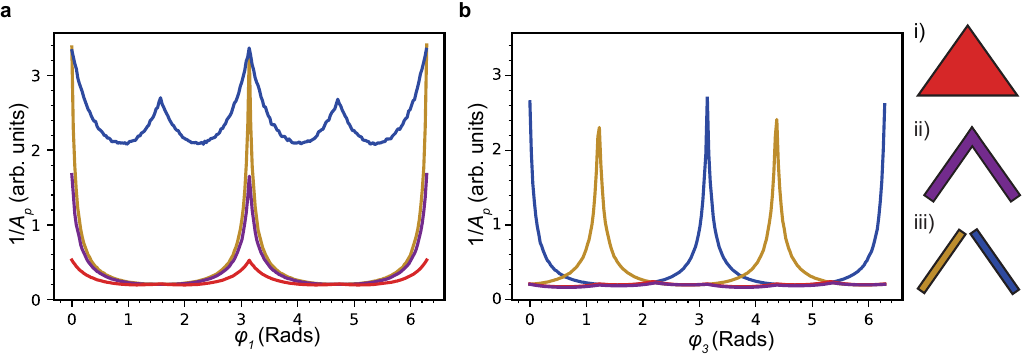}
    \caption{\textbf{Monte Carlo simulations of the inverse projected area.}
    \textbf{a,b}, Calculated inverse projected area $1/A_p$ as a function of rotation angle for $\varphi_1$ (\textbf{a}) and $\varphi_3$ (\textbf{b}) for the three transport geometries considered in the main text: (i) solid triangular prism (red), (ii) connected finite-thickness facets (purple), and (iii) two independent facets (blue/gold). Insets schematically illustrate the corresponding cross-sections of the model geometries.
    }
    \label{fig:MC}
\end{figure}

\clearpage

\section{Tracking UCF Features}

To quantify the angular evolution of the conductance fluctuations, individual interference features were manually tracked across the angle-resolved conductance maps shown in the main text. The extracted trajectories were subsequently used for the normalized inverse projected-area analysis presented in Fig.~2e of the main text.

\begin{figure}[h!]
    \centering
    \includegraphics[width=0.95\linewidth]{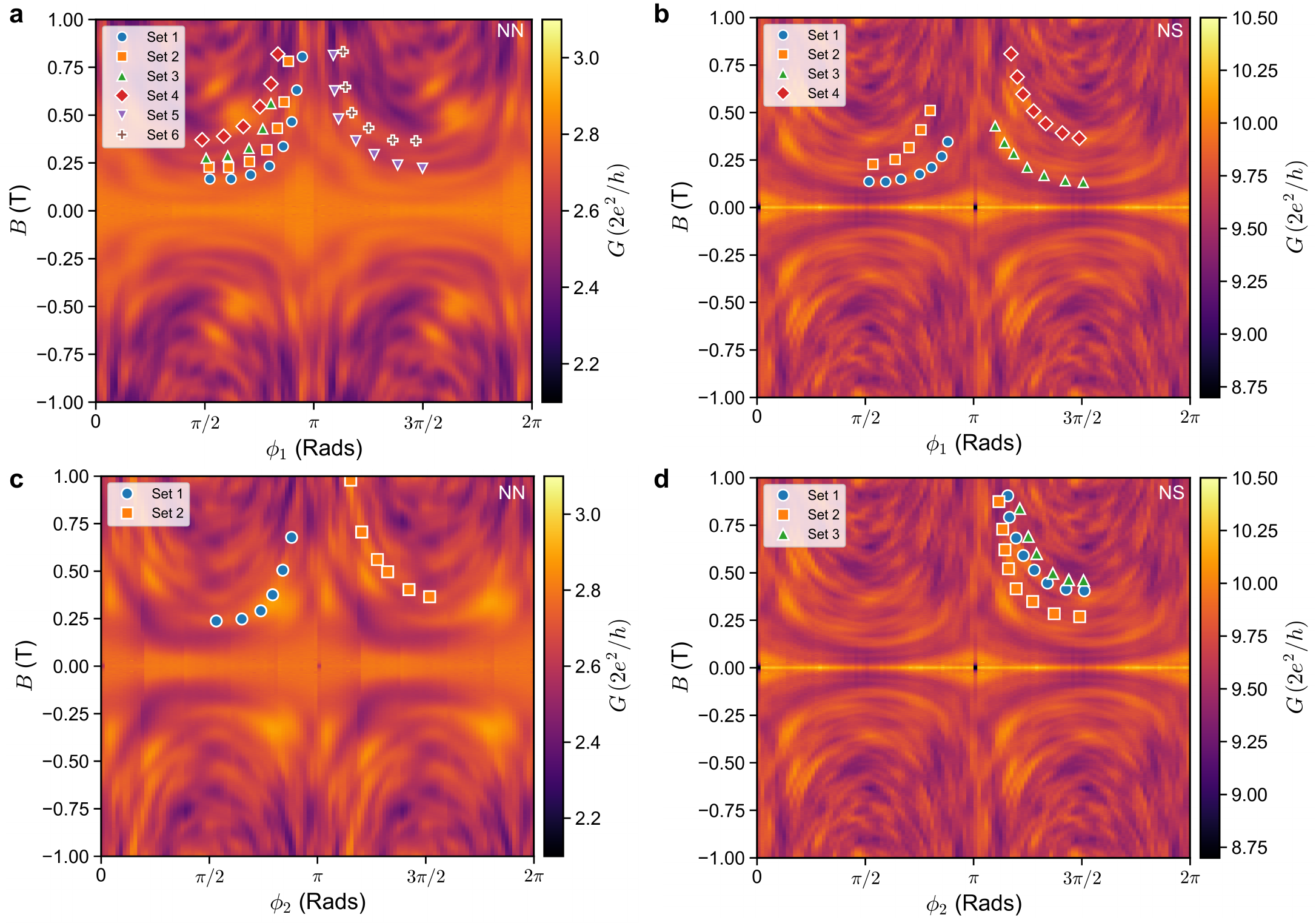}
    \caption{
\textbf{Extracted conductance fluctuation trajectories on angle-resolved conductance maps.}
Color maps of the conductance $G$ as a function of magnetic-field magnitude $B$ and magnetic-field rotation angle $\phi_\mathrm{1},\phi_\mathrm{2}$ for the NN and NS1 device. Overlaid markers indicate manually tracked trajectories corresponding to conductance-fluctuation extrema in the maps. Different marker styles distinguish independently extracted trajectory families.
}
    \label{fig:tracking}
\end{figure}

Figure~\ref{fig:tracking} shows representative examples of the extracted conductance-fluctuation trajectories for both NN and NS devices during rotations in F1 and F2. Distinct fluctuation extrema can be followed continuously as a function of magnetic-field orientation and evolve towards higher magnetic fields as the field approaches the facet plane. Multiple independent trajectory families were extracted for each dataset in order to sample different interference paths contributing to the conductance fluctuations.

The trajectories were extracted using a web-based digitizer tool, where points were manually selected directly from the conductance maps and exported as CSV coordinate files for subsequent plotting and analysis. Each trajectory was normalized to its lowest-field point prior to comparison with the projected-area models discussed in the main text.

